\documentclass[
  aps,
  prl,
  reprint,
  showpacs,
  groupedaddress,
  amsmath,
  amssymb,
  floatfix]{revtex4-1}
  
\usepackage{graphicx,epsfig,dcolumn,multirow}

\usepackage{color}

\newcolumntype{d}[1]{D{.}{.}{#1}}

\RequirePackage{xspace}

\newcommand{\dis}[1]{\begin{equation}\begin{split}#1\end{split}\end{equation}}
\newcommand{\be}{\begin{equation}}
\newcommand{\ee}{\end{equation}}
\def\bea{\begin{eqnarray}}
\def\eea{\end{eqnarray}}

\newcommand{\eq}[1]{Eq.~(\ref{#1})}
\newcommand{\eqs}[1]{Eqs.~(\ref{#1})}

\newcommand{\VEV}[1]{\langle #1 \rangle}

\newcommand{\tev}{\,\textrm{TeV}}
\newcommand{\gev}{\,\textrm{GeV}}

\def\tb{\tan\beta}

\begin{document}
%\draft

\title{\Large\bf 
%The amendments of the focus point scenario
%Deflected 
Shifted focus point 
of the Higgs mass parameter 
\\from the minimal mixed mediation of SUSY breaking 
}

\author{Bumseok Kyae\footnote{email: bkyae@pusan.ac.kr}
}
\affiliation{
Department of Physics, Pusan National University, Busan 609-735, Korea
}

\begin{abstract}

We employ both the minimal gravity- and the minimal gauge mediations of supersymmetry breaking at the grand unified theory (GUT) scale in a single supergravity framework, 
assuming the gaugino masses are generated 
dominantly by the minimal gauge mediation effects.   
In such a ``minimal mixed mediation model,''  
a ``focus point'' of the soft Higgs mass parameter, $m_{h_u}^2$ emerges 
at $3$-$4 \tev$ energy scale, 
which is exactly the stop mass scale needed for explaining the $126 \gev$ Higgs boson mass 
without the ``$A$-term'' at the three loop level. 
As a result, $m_{h_u}^2$ 
%in the minimal supersymmetric standard model 
can be quite insensitive to various trial stop masses at low energy, 
reducing the fine-tuning measures to be much smaller than $100$ 
even for a $3$-$4 \tev$ low energy stop mass and   
$-0.5 < A_t/m_0\lesssim +0.1$ at the GUT scale. 
%The $\mu$ parameter is smaller than $600 \gev$. 
The gluino mass is predicted to be about $1.7 \tev$, 
which could readily be tested at LHC run2.

%minimal SUGRA assisted with the gauge mediated SUSY breaking effects at the GUT scale

\end{abstract}



\maketitle

%%%%%%%%%%%%%%%%%%%%%%%%%%%%%%%%%%%%%%%%%%%%%%%%%%%%%%%%%%%%%%%%%%%%%%%%%%%%%%%%%
%%%%%%%%%%%%%%%%%%%%%%%%%%%%%%%%%%%%%%%%%%%%%%%%%%%%%%%%%%%%%%%%%%%%%%%%%%%%%%%%%

%time to ask naturalness of SUSY even with a heavy stop mass.

%$\bullet$ SUSY, gauge coupling unification, GUT

%$\bullet$ vanishing cosmological constant, SUSY breaking in SUGRA, (SUSY breaking mediations)

%$M_G$ means the grand unified theory (GUT) scale ($\approx 2\times 10^{16}~{\rm GeV}$) adopted as a cut-off of the model.     
%For simple expressions, here we assumed that the ``$A$-term'' coefficient corresponding to the top quark Yukawa coupling,  
%$A_t$ dominates over $\mu \cdot{\rm cot}\beta$, where $\mu$ stands for the ``$\mu$-term'' coefficient in the MSSM superpotential.
%
%$\Lambda$ means a cutoff scale. A messenger scale of SUSY breaking is usually adopted for it. 

Although the standard model (SM) has been extremely successful 
in the experimental side, 
it doesn't provide reasonable answers to some theoretical puzzles 
such as the naturalness of the electroweak (EW) scale 
and the Higgs boson mass.  
The main motivation of the low energy supersymmetry (SUSY) was to resolve 
the naturalness problem associated with the EW phase transition 
raised in the SM, since SUSY can protect the small Higgs mass 
against large quantum corrections \cite{book,PR1984}. 
%With this respect of SUSY, 
Because of it, the minimal supersymmetric standard model (MSSM) 
has been believed the most promising theory beyond the SM,  
guiding the SM to a grand unified theory (GUT) or string theory.
However, any evidence of the low energy SUSY has not been observed yet 
at the large hadron collider (LHC):  
the mass bounds on the SUSY particles have gradually increased, and 
now they seem to start threatening the traditional status of SUSY 
as a prominent solution to such a naturalness problem of the SM. 
% of the EW phase transition.
%It would be timely to discuss again the naturalness problem

Actually, a barometer of the naturalness of the MSSM is the mass of the superpartner of the top quark (``stop'').  
Due to the large top quark Yukawa coupling ($y_t$), the top and stop make the dominant contributions to the radiative physical Higgs mass squared and also 
the renormalization of a soft mass squared of the Higgs ($m_{h_u}^2$) in the MSSM. 
The renormalization effect on $m_{h_u}^2$ would linearly be sensitive to  
the stop mass squared $\widetilde{m}_t^2$ \cite{book},  
\bea \label{renormHiggs}
%\label{physHiggs}
%&&\Delta m_H^2\approx 
%\frac{3y_t^4}{4\pi^2}{\rm sin}^4\beta v_h^2 ~{\rm log}\left(\frac{\widetilde{m}_t^2}{m_t^2}\right) + \cdots , 
%%\frac{A_t^2}{\widetilde{m}_t^2}\left(1-\frac{1}{12}\frac{A_t^2}{\widetilde{m}_t^2}\right)\right] , 
%\\  
%&&  
%~~
\Delta m_{h_u}^2\approx \frac{3y_t^2}{8\pi^2}\widetilde{m}_t^2 
~{\rm log}\left(
\frac{\widetilde{m}_t^2}{\Lambda^2}\right) +\cdots ,  
%\left[1 + \frac12\frac{A_t^2}{\widetilde{m}_t^2}\right] ,
\eea
%where $m_t$ ($\widetilde{m}_t$) denotes the top quark (stop) mass, and $v_h$ 
%is the vacuum expectation value (VEV) of the Higgs boson, $v_h\equiv\sqrt{\VEV{h_u}^2+\VEV{h_d}^2}\approx 174~{\rm GeV}$  with $\tan\beta\equiv\VEV{h_u}/\VEV{h_d}$. 
while it depends just logarithmically on a ultraviolet (UV) cutoff $\Lambda$.  
Since the Higgs mass parameters, $m_{h_u}^2$ and $m_{h_d}^2$ are related to 
the the $Z$ boson mass $m_Z$ 
together with 
%the ``$\mu$-term'' coefficient ($\mu$) in the MSSM, 
the ``Higgsinos'' (superpartners of the Higgs boson) mass, $\mu$ \cite{book}, 
\dis{ \label{m_Z}
\frac12 m_Z^2=\frac{m_{h_d}^2-m_{h_u}^2{\rm tan}^2\beta}{{\rm tan}^2\beta-1}
-|\mu|^2 , 
}
%where $m_Z^2$ denotes the $Z$ boson mass and $\mu$ is the ``$\mu$-term'' coefficient in the MSSM superpotential. 
$\{m_{h_u}^2, m_{h_d}^2, |\mu|^2\}$ should be finely tuned 
to yield $m_Z^2=(91 \gev)^2$ for a given $\tb$ [$\equiv\VEV{h_u}/\VEV{h_d}$, 
ratio of the vacuum expectation values (VEVs) 
of the two MSSM Higgs fields], 
if they are excessively large.  
According to the recent analysis based on the three-loop calculations, 
the stop mass required for explaining the $126 \gev$ Higgs boson mass \cite{LHCHiggs} 
without any other helps is about $3$-$4 \tev$ \cite{3-loop}.  
Thus, a fine-tuning of order $10^{-3}$ or smaller looks unavoidable
in the MSSM for $\Lambda\sim 10^{16} \gev$. 

In order to more clearly see the UV dependence of $m_{h_u}^2$ and 
properly discuss this ``little hierarchy problem'', 
however, one should suppose a specific UV model  
and analyze its resulting full renormalization group (RG) equations. 
If the UV model is simple enough, addressing this problem successfully with SUSY,
the low energy SUSY could still be regarded as an attractive solution 
to the naturalness problem.  

One nice idea is the ``focus point (FP) scenario'' \cite{FMM1}. 
This scenario is based on the minimal gravity mediation (mGrM) of SUSY breaking. 
So the soft mass squareds 
such as $m_{h_{u,d}}^2$ and those of the left handed (LH) 
and right handed (RH) stops, ($m_{q_3}^2, m_{u_3^c}^2$) 
as well as the gaugino (superpartners of the gauge fields) masses 
$M_a$ ($a=3,2,1$) 
are given to be {\it universal} at the GUT scale, $m_{h_u}^2=m_{h_d}^2=m_{q_3}^2=m_{u_3^c}^2=\cdots\equiv m_0^2$ 
and $M_3=M_2=M_1\equiv m_{1/2}$. 
As pointed out in \cite{FMM1}, 
if the holomorphic soft SUSY breaking terms (``$A$-terms'') 
in the scalar potential
are zero at the GUT scale and 
the unified gaugino mass $m_{1/2}$ 
is just a few hundred GeV, 
$m_{h_u}^2$ converges to a small negative value 
%[$|m_{h_u}^2|\ll (1 \tev)^2$] 
around the $Z$ boson mass scale in this setup, 
{\it regardless of its initial values given by $m_0^2$ at the GUT scale} \cite{FMM1}:
a FP of $m_{h_u}^2$ appears around the $m_Z$ scale. 
In the RG solution of $m_{h_u}^2$ at the $m_Z$ scale, namely,  
\dis{ \label{RGsol}
m_{h_u}^2(Q=m_Z) = C_{s} m_0^2-C_{g} m_{1/2}^2, 
}
where the dimensionless numbers $C_{s}$, $C_{g}$ ($>0$) can numerically be estimated  
using RG equations, 
$C_{s}$ happens to be quite small 
with the above universal soft masses, 
and the EW symmetry is broken dominantly by the $C_g$ term.  
%if RG equations  are valid down to about $m_Z$ scale. 
On the other hand, stop masses are quite sensitive to $m_0^2$. 
As a result, $m_Z^2$ could remain small enough 
even with a relatively heavy stop mass in the FP scenario 
in contrast to the naive expectation from \eq{renormHiggs}. 
  
However, the experimental bound on the gluino (superpartner of the gluon) mass $M_3$ 
has already exceeded $1.3 \tev$ \cite{gluinomass}. 
As expected from Eqs.~(\ref{m_Z}) and (\ref{RGsol}), 
a too large $m_{1/2}$ needed for $M_3>1.3 \tev$ at low energy 
%leads to a too large negative $m_{h_u}^2$ at the EW scale \cite{KS}. 
would require a fine-tuned large $|\mu|$ for $m_Z$ of $91 \gev$ 
particularly for relatively light stop mass ($\lesssim 1 \tev$) cases. 
When the stop mass is around $3$-$4 \tev$, 
the stop should decouple from the RG equations below $3$-$4 \tev$, 
which makes $C_{s}$ {\it sizable} in \eq{RGsol} \cite{KS}. 
Then, a much larger $m_{1/2}$ is necessary for EW symmetry breaking.  
Since the RG running interval between $3$-$4 \tev$ and $m_Z$ scale, 
to which modified RG equations should be applied, is too large,  
the FP behavior is seriously spoiled    
with such heavy SUSY particles. 
%which results in quite large fine-tunings. 

The best way to rescue the FP idea is to somehow shift the FP upto the stop decoupling scale \cite{KS}: 
$C_s$ needs to be made small enough before stops are decoupled.   
Then $m_{h_u}^2$ at the $m_Z$ scale can be estimated using the Coleman-Weinberg potential \cite{book,CQW}: 
\bea \label{RGsm}
&&m_{h_u}^2(m_Z)\approx m_{h_u}^2(\Lambda_T) + 
\frac{3|y_t|^2}{16\pi^2}\bigg[{m}_{q_3}^2\left\{{\rm log}\frac{{m}_{q_3}^2}{\Lambda_T^2}-1\right\}
\nonumber \\
&&\qquad\quad + {m}_{u_3^c}^2\left\{{\rm log}\frac{{m}_{u_3^c}^2}{\Lambda_T^2}-1\right\}
-m_t^2\left\{{\rm log}\frac{m_t^2}{\Lambda_T^2}-1\right\}\bigg]
\nonumber \\ 
&&\qquad \approx m_{h_u}^2(\Lambda_T) - \frac{3|y_t|^2}{16\pi^2}
\left({m}_{q_3}^2+m_{u_3^c}^2\right)\bigg|_{\Lambda_T} , 
\eea
where the cutoff $\Lambda_T$ is set to the stop decoupling scale [$\approx (m_{q_3}m_{u_3^c})^{1/2}$], and the top quark mass ($m_t$) contributions are relatively suppressed. 
%where the stops are decoupled, and so $m_{h_u}^2|_{\Lambda_T} = m_{h_u}^2(t_T)$.  
%Here we take $m_{u^c_3}^2\approx m_{q_3}^2\equiv\widetilde{m}_t^2$ for a simple estimation.  
Since the $m_0^2$ dependence of stop masses would be loop-suppressed,   
$m_{h_u}^2$ needs to be well-focused around $\Lambda_T$. 
Due to the additional negative contribution to $m_{h_u}^2(m_Z)$ below $\Lambda_T$, 
a small positive $m_{h_u}^2(\Lambda_T)$ would be more desirable.

In order to push up the FP to the desired stop mass scale $3$-$4 \tev$, 
in this letter we suggest to combine 
the two representative SUSY breaking mediation scenarios, 
the mGrM and the minimal gauge mediation (mGgM) 
in a single supergravity (SUGRA) framework 
with a {\it common} SUSY breaking source.  
We will call it ``minimal mixed mediation.''

%********************************************
%
%In fact one can easily obtain such a universal soft mass  
%for Higgs and the third generations of MSSM super  
%by employing a simple gravity mediation of SUSY breaking. 
%In this case, the FP of $m_{h_u}^2$ appears below $500 \gev$ XXX, 
%but accompanied with a too light stop mass.  
% 
%**********************************************

The chiral SUGRA Lagrangian is basically described 
in terms of the K${\rm\ddot{a}}$hler potential $K$, 
superpotential $W$, and gauge kinetic function. 
First, let us consider the minimal K${\rm\ddot{a}}$hler potential, 
and a superpotential where the observable and hidden sectors are separated 
as in the ordinary mGrM \cite{book}:  
\dis{ \label{KSpot} 
K=\sum_{i,a} |z_i|^2+|\phi_a|^2 ~, \quad
W=W_H(z_i)+W_O(\phi_a)
}
where $z_i$ [$\phi_a$] denotes fields in the hidden [observable] sector, 
carrying hidden [SM or GUT] gauge quantum numbers. 
The kinetic terms of $z_i$ and $\phi_a$, thus, take the canonical form. 
We assume non-zero VEVs for $z_i$s \cite{PR1984}: 
\dis{ \label{vev}
\langle z_i\rangle=b_iM_P , ~ 
\langle\partial_{z_i} W_H\rangle=a_i^*mM_P , ~ 
\langle W_H\rangle=mM_P^2 ,
}
where $a_i$ and $b_i$ are dimensionless numbers, 
while $M_P$ ($\approx 2.4\times 10^{18} \gev$) denotes the reduced Planck mass. 
Then, $\langle W_H\rangle$ or $m$ gives the gravitino mass, 
$m_{3/2}=e^{K/2M_P}\langle W\rangle/M_P^2=e^{|b_i|^2/2}m$.  
The soft terms can read from the scalar potential of SUGRA:
\dis{ \label{sugraPot}
V_F=e^{\frac{K}{M_P^2}}\left[\left|F_{z_i}\right|^2
+\left|F_{\phi_a}\right|^2-\frac{3}{M_P^2}|W|^2\right]
}
where the ``$F$-terms,'' 
$F_i$ ($=D_iW=\partial_iW+\partial_iK ~W/M_P^2$) are given by 
\begin{eqnarray}
&&F_{z_i}=\frac{\partial W_H}{\partial z_i}+z_i^*\frac{W}{M_P^2} 
=M_P\left[\left(a_i^*+b_i^*\right)m+b_i^*\frac{W_O}{M_P^2}\right] ,
\nonumber \\  \label{F_a}
&&F_{\phi_a}=\frac{\partial W_O}{\partial \phi_a}+\phi_a^*\frac{W}{M_P^2}
=\frac{\partial W_O}{\partial \phi_a}+\phi_a^*\left(m+\frac{W_O}{M_P^2}\right) . 
\end{eqnarray}
%\dis{
%F_{z_i}=M\left[\left(a_i^*+b_i^*\right)m+b_i^*\frac{W_O}{M_P^2}\right] ,~~
%F_{\phi_a}=\frac{\partial W_O}{\partial \phi_a}+\phi_a^*\left(m+\frac{W_O}{M_P^2}\right) %. 
%}
The vanishing cosmological constant (C.C.) requires a fine-tuning 
between $\langle F_{z_i}\rangle$ and $\langle W_H\rangle$, i.e. from \eq{sugraPot} 
$\sum_i\langle |F_{z_i}|^2\rangle=3|\langle W_H\rangle|^2/M_P^2$, or 
$\sum_{i}\left|a_i+b_i\right|^2=3$.
Neglecting the non-renormalizable terms suppressed with $1/M_P^2$, 
\eq{sugraPot} is rewritten as \cite{PR1984} 
\dis{ \label{scalarPot}
&\qquad\quad V_F \approx \left|\partial_{\phi_a}\widetilde{W}_O\right|^2
+m_{0}^2|\phi_a|^2
\\
&+m_{0}\left[\phi_a\partial_{\phi_a}\widetilde{W}_O
+(A_\Sigma-3)\widetilde{W}_O+{\rm h.c.}\right] .
}
where $A_\Sigma$ is defined as $A_\Sigma\equiv \sum_ib_i^*(a_i+b_i)$. 
$m_0$ is identified with the gravitino mass $m_{3/2}$ ($=e^{|b_i|^2/2}m$) and  
$\widetilde{W}_O$ ($\equiv e^{|b_i|^2/2}W_O$) 
denotes the rescaled $W_0$. 
From now on, we will drop out the ``tilde'' for a simple notation.
The first term of \eq{scalarPot} corresponds to 
the $F$-term potential in global SUSY, the second term is 
the universal soft mass term, and  
the remaining terms are 
%soft SUSY breaking 
$A$-terms.    
The {\it universal} $A$-parameter here ($\equiv A_0=A_t$) 
does not include Yukawa coupling constants, 
but it is proportional to $m_0$.
If there is no quadratic term or higher powers of $\phi_a$ in $W_O$, 
one can get negative (positive) $A$-terms with $A_\Sigma <2$ ($A_\Sigma >2$). 
With the vanishing C.C. condition, the universal soft mass parameter,  
$m_0$ ($=e^{\langle K\rangle/2M_P^2}\langle W_H\rangle/M_P^2$)
can be recast to 
$e^{\langle K\rangle/2M_P^2}\left(\sum_i|\langle F_{z_i}\rangle|^2\right)^{1/2}/\sqrt{3}M_P$, 
which is the conventional form in the mGrM scenario. 
%Such a universal soft parameter could be provided from the minimal form of the K${\rm\ddot{a}}$hler potential 

%****************************************
%
%In order to push up the scale of the FP, one could utilize a sizable Dirac neutrino mass term $y_Nl_3h_uN^c$ as discussed in Ref.~\cite{KS}: 
%Such a coupling ($y_N\approx 1.0$ XXX) makes the top quark Yukawa coupling much larger 
%through the RG evolution at higher energies, and can eventually move the FP to a higher energy scale as seen in Eq.~(\ref{}).  
%However, there is no guiding principle to set $y_N$ to the precise value required for a FP emerging at the stop decoupling scale. 
%For instance, introduction of SO(10) or flipped SU(5) GUT can bind $y_N$ and $y_t$ to a single Yukawa coupling at the GUT scale.  
%$y_N$ determined in such a way ($\lesssim 0.3$) is, however, too small to uplift the FP sufficiently.    

%In this letter, we suggest another way to push up the FP scale 
%such that it appears around the desired stop decoupling scale ($3$-$4 \tev$): 
%on top of the universal soft masses induced by the mGrM, 
%we add the mGgM effects at the GUT scale. 
% only a fine-tuning for the vanishing C.C. 
%    
%****************************************    
    
Next, let us introduce one pair of messenger superfields $\{{\bf 5},\overline{\bf 5}\}$, 
which are the SU(5) fundamental representations, protecting the gauge coupling unification. 
Through their coupling with a SUSY breaking source $S$, which is an MSSM  singlet superfield,  
\dis{ \label{Wm}
W_m=y_SS{\bf 5}\overline{\bf 5} ,
}
the soft masses of the MSSM gauginos and scalar superpartners are generated 
at one- and two-loop levels, respectively \cite{book}:
\dis{ \label{GGsoft}
M_a= %N_5
\frac{g_a^2}{16\pi^2}\frac{\langle F_S\rangle}{\langle S\rangle} ,~
m_i^2=2 %N_5
\sum_{a=1}^{3}
\left[\frac{g_a^2}{16\pi^2}\frac{\langle F_S\rangle}{\langle S\rangle}\right]^2C_a(i) 
}
where $C_a(i)$ is the quadratic Casimir %group theory 
invariant for a superfield $i$, $(T^aT^a)_i^j=C_a(i)\delta_i^j$, and   
$g_a$ ($a=3,2,1$) denotes the MSSM gauge coupling constants. 
$\langle S\rangle$ and $\langle F_S\rangle$ are VEVs of the scalar and $F$-term components of the superfield $S$. 
Note that $M_a$ and $m_i^2$ are almost independent of $y_S$ 
only if $\langle F_S\rangle\lesssim y_S^2\langle S\rangle$ \cite{book}. 
However, such mGgM effects would appear below the messenger scale, 
$y_S\langle S\rangle$.  
Here we assume that $\langle S\rangle$ has the same magnitude as
the VEV of the SU(5) breaking Higgs $v_G$: 
% identified with  
$\langle {\bf 24}_H\rangle=v_G\times {\rm diag.}(2,2,2;-3,-3)/\sqrt{60}$. 
It is possible if a GUT breaking mechanism causes $\langle S\rangle$ \cite{on-going}. 
Actually, the masses of ``$X$'' and ``$Y$'' gauge bosons induced by 
%the VEV of the GUT breaking Higgs,   
$\langle {\bf 24}_H\rangle$, 
$M_X^2=M_Y^2=\frac{5}{24}g_G^2v_G^2$ \cite{GUT}, 
where $g_G$ is the unified gauge coupling constant,  
can be identified with the MSSM gauge coupling unification scale, 
because the unified gauge interactions would become active 
above the $M_{X,Y}$ scale. 

%As pointed out in Ref.~\cite{KimKim}, if $S$ itself is a GUT breaking Higgs superfield, 
%$F_S$ directly couples to the visible fields through gauge interactions,  
%and tachyonic soft squared masses are generated at the GUT scale. 
%In this letter, we regard $S$ as an MSSM singlet as in the ordinary mGgM.
%Instead we suppose that a GUT breaking Higgs causes $\langl S\rangle$ at the minimum of 
%the $F$- or $D$-term potential \cite{on-going}, and  
%
%
%We set $\langle F_S\rangle=m_0M_P$ as in the mGrM. 
%It is possible to hold $\langle F_S\rangle\sim m_0M_P$ (rather than $m_0v_G$) 
%under a mechanism to yield $\langle S\rangle=v_G$:   

In addition to \eq{KSpot}, the K${\rm\ddot{a}}$hler potential 
(and hidden local symmetries we don't specify here) 
can permit     
\dis{ \label{Extpot}
K\supset f(z)S +{\rm h.c.} ,
}      
where $f(z)$ denotes a {\it holomorphic} monomial of hidden sector fields $z_i$s 
with VEVs of order $M_P$ in \eq{vev}, 
and so it is of order ${\cal O}(M_P)$. 
Their kinetic terms still remain canonical.  
The U(1)$_R$ symmetry forbids $M_Pf(z)S$ in the superpotential.
%If $\langle\partial_SW\rangle$ is relatively suppressed, 
Then, the resulting $\langle F_S\rangle$ can be  
\dis{
\langle F_S\rangle\approx m\left[\langle f(z)\rangle+\langle S^*\rangle\right] 
}
by including the SUGRA corrections with $\langle W_H\rangle=mM_P^2$. 
Thus, the VEV of $F_S$ is of order ${\cal O}(mM_P)$ like $F_{z_i}$ in \eq{F_a}. 
They should be fine-tuned for the vanishing C.C.:  
a precise determination of $\langle F_S\rangle$ is indeed 
associated with the C.C. problem.
%Since $\langle F_{{\bf 24}^c}\rangle \ll \langle F_S\rangle$, the gauge messenger effects by $F_{{\bf 24}^c}$ 
%are relatively suppressed. 
%
%Note that $\langle F_S\rangle$ 
%%and $\langle F_{{\bf 24}^c}\rangle$ 
%doesn't make contributions to $a_i^*$ of \eq{vev}.  
%because $\langle \partial_SW\rangle=\langle\partial_{{\bf 24}^c}W\rangle=0$.
%Assuming $A_\Sigma\lesssim 1$ ($\gtrsim 1$) from $\{F_{z_i}\}$, thus, 
%from $\{F_z,F_{\bar{z}},F_{z^c},F_{\bar{z}^c}\}$, thus, 
%one can get small negative (positive)  $A$-terms at the GUT scale from \eq{scalarPot}.   
%
Here we set $\langle F_S\rangle=m_0M_P$. 
%as in the mGrM.
%
$F_{\phi_a}$ is still given by \eq{F_a}, 
which induces the universal soft mass terms at tree level.

%We will comment later on how to realize $\langle S\rangle=v_G$ 
%and $\langle F_S\rangle\sim m_0M_P$. 
%
Thus, the typical size of mGgM effects is estimated as
\dis{ \label{GM}
\frac{\langle F_S\rangle}{16\pi^2\langle S\rangle}=\frac{m_0 M_P}{16\pi^2M_X}\sqrt{\frac{5}{24}}g_{G}
\approx 0.36 \times m_0 . ~~~ 
}
Here we set the unified gauge coupling at the GUT scale 
[$\approx (1.3\pm 0.4)\times 10^{16} \gev$] to $g_{G}^2/4\pi\approx 1/26$ 
due to relatively heavy colored superpartners ($\gtrsim 3 \tev$). 
%It means $\langle S\rangle=v_G\approx \cred{4}\times 10^{16} \gev$.  
Even for $|y_S|\ll 1$, we will keep this value,   
%for $\langle S\rangle$. 
since it is fixed by a UV model.

The fact that the mGgM effects by \eq{GGsoft} are proportional to $m_0$ or $m_0^2$
are important. 
Moreover, $A$-terms from \eq{scalarPot} are also proportional to $m_0$.  
%As seen in the semi-analytic solutions of Ref.~\cite{KS}, 
In this setup, thus,    
an (extrapolated) FP of $m_{h_u}^2$ must still exist
at a higher energy scale \cite{on-going}.    
%i.e. $m_{h_u}^2(Q_T)/m_0^2\approx 0$ \cite{on-going}.  
%only if the tree level gaugino mass is suppressed. 
%since they all are proportional to $m_0^2$. 
%$C_g$ in \eq{RGsol} is absorbed in $C_s$, 
%making it small again: 
As $C_g$ is converted to a member of $C_s$ in \eq{RGsol}, 
the naturalness of $m_{h_u}^2$ and $m_Z^2$ becomes gradually improved, 
making $C_s$ smaller and smaller, 
until the FP reaches the stop decoupling scale.

For $|y_S|\lesssim 1$ in \eq{Wm}, 
the messenger scale $Q_M$ drops down below $M_{X,Y}$.  
Since $X$ and $Y$ gauge sectors have already been decoupled below the messenger scale,
the soft masses generated by the mGgM in \eq{GGsoft} 
become {\it non-universal} for $Q_M<M_{X,Y}$. 
Of course, the beta function coefficients of the MSSM fields should be modified 
above the scale of $y_S\langle S\rangle$ by  
the messenger fields $\{{\bf 5}, \overline{\bf 5}\}$. 
Thus, the RG equations of the MSSM gauge couplings and gaugino masses are 
\dis{ \label{RGgauge}
8\pi^2\frac{dg_a^2}{dt}=b_ag_a^4 , ~~8\pi^2\frac{dM_a}{dt}=b_ag_a^2M_a ,
}
where $t\equiv{\rm log}[Q/\gev]$, and $b_a=(-2,2,\frac{38}{5})$ for $Q>Q_M$  
while $b_a=(-3,1,\frac{33}{5})$ for $Q<Q_M$. 
For the RG equations of the Yukawa couplings of 
the third generation of quarks and leptons $(y_t,y_b,y_\tau)$ and other soft parameters, 
refer to Appendix of Ref.~\cite{KS}.

The boundary conditions at the GUT scale are given by the universal form 
as seen in \eq{scalarPot}. 
Unlike the case of the mGrM, we have {\it additional} non-universal contributions 
by \eq{GGsoft}. 
They should be imposed at a given messenger scale,  
and so affect the RG evolutions of MSSM parameters for $Q\leq Q_M$.
To see clearly how the original FP scenario is modified 
by the additional mGgM effects, 
in this letter we don't consider the superheavy RH neutrinos 
in the RG analysis as in \cite{FMM1}, 
assuming their couplings are small enough, 
even if they are helpful for improving the naturalness \cite{KS,RHneu}. 

We also suppose that the gaugino masses from the mGrM  
are relatively suppressed. 
In fact, the gaugino mass term in SUGRA is associated with the first derivative of the gauge kinetic function \cite{PR1984}, 
and so a constant gauge kinetic function at tree level 
($= \delta_{ab}$) can realize it.  
In fact, it is the simplest case, 
yielding the canonical gauge kinetic terms in the Lagrangian.   
%[$\supset\frac{-1}{4}{\rm Re}f_{ab}F_{\mu\nu}^aF^{b\mu\nu}=\frac{-1}{4}F_{\mu\nu}^aF^{a\mu\nu}$],    
Accordingly, the gaugino masses by \eq{GGsoft} dominates over them in this case. 
Then \eqs{GGsoft}, (\ref{GM}), and (\ref{RGgauge}) admit a simple analytic expression 
for the gaugino masses at the stop mass scale: 
\dis{ \label{M_a}
M_a(t_T)\approx 0.36\times m_0\times g_a^2(t_T) ,
} 
It does {\it not} depend on messenger scales. 
 
%%%%%%%%%%%%%%%%%%%%%%%%%%%%%%%%%%%%%%%%%%%%%%%%%%%%%%%%%%%
%%%%%%%%%%%%%%%%%%%%%%%%%%%%%%%%%%%%%%%%%%%%%%%%%%%%%%%%%%%
%
%
\begin{figure}
\begin{center}
\includegraphics[width=0.9\linewidth]{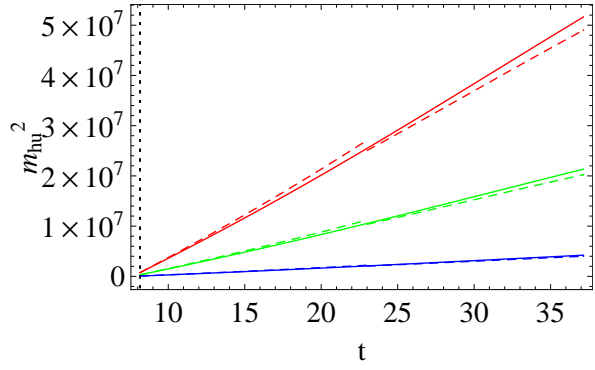}
\end{center}
\caption{RG evolutions of $m_{h_u}^2$ 
with $t$ [$\equiv {\rm log}(Q/\gev)]$
for $m_0^2=(7 \tev)^2$ [Red], $(4.5 \tev)^2$ [Green], 
and $(2 \tev)^2$ [Blue] when $A_t=-0.2 ~m_0$ and $\tb=50$. 
The tilted straight [dotted] lines correspond to the case of 
$t_M\approx 37$ (or $Q_M\approx 1.3\times 10^{16} \gev$, ``{\rm Case A}'') 
[$t_M\approx 23$ (or $Q_M=1.0\times 10^{10} \gev$, ``{\rm Case B}'')].
%The discontinuities should arise at the messenger scales. 
%The FP is not affected by messenger scales. 
The vertical dotted line at $t=t_T\approx 8.2$ ($Q_T=3.5 \tev$) indicates 
the desired stop decoupling scale. 
The discontinuities of $m_{h_u}^2(t)$ should appear at the messenger scales.
%when the mGrM of SUSY breaking is combined with the GUT scale gauge mediation, the FP of $m_{h_u}^2$ is shifted to $t=t_T$, 
%which is the desired stop mass scale for explaining $126 \gev$ Higgs mass without $A_t$. 
%The contribution of $A_t$ to the Higgs boson mass is still negligible.   
%
%
%{\bf (a)} Additional SUSY mass of the superfields $N$,
%$\overline{N}$ generated by the Higgs VEVs. {\bf (b)} A
%contribution to the radiatively induced effective Higgs potential.
%The trilinear scalar couplings in {\bf (a)} and {\bf (b)} come
%from the cross term of $|\partial{W}/\partial{S}|^2$.
}
\label{fig:1}
\end{figure}
%
%
%%%%%%%%%%%%%%%%%%%%%%%%%%%%%%%%%%%%%%%%%%%%%%%%%%%%%%%%%%%
%%%%%%%%%%%%%%%%%%%%%%%%%%%%%%%%%%%%%%%%%%%%%%%%%%%%%%%%%%%

%Addition of such a non-universal boundary conditions for soft mass parameters at a  messenger scale on top of the universal spectrum by the mGrM
%
%Employing both the mGrM and mGgM at the GUT scale 
%turns out to move  
%the original FP of the mGrM to $3$-$4 \tev$, 
%which is exactly the desired stop decoupling scale for explaining the $126 \gev$ Higgs boson mass without $A_t$. See Fig.~\ref{fig:1}. 
Fig.~\ref{fig:1} displays RG evolutions of $m_{h_u}^2$ for various trial $m_0^2$s. 
%[$=(7 \tev)^2$, $(4.5 \tev)^2$, $(2 \tev)^2$].  
%when $A_t=-0.5~m_0$ and $\tb=50$. 
%
The straight [dotted] lines correspond to the case of 
$t_M\approx 37$ (or $Q_M\approx 1.3\times 10^{16} \gev$, ``Case A'') 
[$t_M\approx 23$ (or $Q_M=1.0\times 10^{10} \gev$, ``Case B'' )].
The discontinuities of the lines by additional boundary conditions arise at the messenger scales. 
%
%The vertical dotted line indicates $t=t_T\approx 8.2$ ($Q_T=3.5 \tev$):
%when the mGrM of SUSY breaking is combined with the GUT scale gauge mediation, the FP of $m_{h_u}^2$ is shifted to $t=t_T$, 
%which is the desired stop mass scale for explaining $126 \gev$ Higgs mass without $A_t$. 
%The contribution of $A_t$ to the Higgs boson mass is still negligible.  
As seen in Fig.~\ref{fig:1}, a FP of $m_{h_u}^2$ appears always at $t=t_T\approx 8.2$ (or $Q_T\approx 3.5 \tev$) {\it regardless of the messenger scales} that we take. 
% 
%
%In fact, the stop mass scales in the both cases are at most about $\{(6 \tev)^2, (4 \tev)^2, (2 \tev)^2\}$ 
%for $m_0^2=\{(6 \tev)^2, (4 \tev), (2 \tev)^2\}$, respectively.
%Below the stop decoupling scales, their RG evolutions should follow modified RG equations. 
%But one can ignore such deviations from our RG evolutions, since the interval between $3.5 \tev$ and the actual stop decoupling scale, $6 \tev$ or $2 \tev$ is too small ($\Delta t<0.6$): 
%its effect on $m_{h_u}^2$ should be smaller than $\frac13$ digits of the vertical axis in Fig.~\ref{fig:1} [one digit is $(\sqrt{2} \tev)^2$], 
%which is hard to destroy the FP behaviors.   
Hence, the wide ranges of UV parameters can yield almost the same values of $m_{h_u}^2$ at low energy.  
Under such a situation, one can guess that $m_0^2\approx (4.5 \tev)^2$ happens to be selected, yielding $3$-$4 \tev$ stop mass, 
and so eventually gets responsible for the $126 \gev$ Higgs mass.

In both cases of Fig.~\ref{fig:1}, the gluino, wino, and bino 
(superpartners of the SM gauge bosons) masses at low energy are 
\dis{ \label{lowM_a}
M_{3,2,1}\approx \{1.7 \tev, ~660 \gev, ~360 \gev\} 
}
for $m_0^2=(4.5 \tev)^2$. 
They are the prediction of this model. They would be testable at LHC run2.
$A_t$ at low energy is about 
%$\{1.0 \tev, 0.5 \tev\}$ 
$1 \tev$ for Case A and B. 
%, respectively. 
Consequently, the contributions of $A_t^2/\widetilde{m}_t^2$ to the radiative Higgs mass 
are smaller than 2.3 $\%$ of those by the stops.

%
%
%The other factors in $\widetilde{m}_i^2$ of \eq{GGsoft} are 
%($i=q_3, u_3^c, d_3^c, h_u, h_d, l_3, e_3^c$) are 
%\dis{
%&2\left[\frac43 g_3^4+\frac34 g_2^4+\frac{1}{60}g_1^4\right]\quad {\rm for}~~q_3 , 
%\\
%&2\left[\frac43 g_3^4+\frac{4}{15}g_1^4\right]\quad {\rm for}~~u_3^c ,
%\\
%&2\left[\frac43 g_3^4+\frac{1}{15}g_1^4\right]\quad {\rm for}~~d_3^c ,
%\\
%&2\left[\frac34 g_2^4+\frac{3}{20}g_1^2\right]\quad {\rm for}~~h_u,~h_d,~l_3 ,
%\\
%&2\left[\frac35 g_1^4\right]\qquad\qquad~ {\rm for}~~e_3^c ,
%}
%
%

%%%%%%%%%%%%%%%%%%%%%%%%%%%%%%%%%%%%%%%%%%%%%%%%%%%%%%%%%%%%%%%%%%
%%%%%%%%%%%%%%%%%%%%%%%%%%%%%%%%%%%%%%%%%%%%%%%%%%%%%%%%%%%%%%%%%%
 %
%\vspace{0.5cm}
%
%
\begin{table}[!h]
\begin{center}
\begin{tabular}
{c|ccc}
\hline\hline
 {\bf Case I} & {\footnotesize $A_t=0$}  & {\footnotesize $\tb=50$}~  & {\footnotesize ${\bf \Delta_{m_0^2}=1}$}  
\\ %\hline
 {\footnotesize ${\bf m_0^2}$} & {\footnotesize $({\bf 5.5} \tev)^2$}  & {\footnotesize $({\bf 4.5} \tev)^2$}  & {\footnotesize $({\bf 3.5} \tev)^2$}  
\\ \hline
 {\footnotesize $m_{q_3}^2(t_T)$} &  {\footnotesize $(4363 \gev)^2$}  & {\footnotesize $(3551 \gev)^2$}  & {\footnotesize $(2744 \gev)^2$} 
\\
 {\footnotesize $m_{u^c_3}^2(t_T)$} &  {\footnotesize $(3789 \gev)^2$}  & {\footnotesize $(3098 \gev)^2$}  & {\footnotesize $(2406 \gev)^2$} 
\\
 {\footnotesize ${\bf m_{h_u}^2(t_T)}$} & {\footnotesize $({\bf 431} \gev)^2$}  & {\footnotesize $({\bf 189} \gev)^2$}  & {\footnotesize $-({\bf 251} \gev)^2$}  
\\
 {\footnotesize $m_{h_d}^2(t_T)$} &  {\footnotesize $(2022 \gev)^2$}  & {\footnotesize $(1512 \gev)^2$}  & {\footnotesize $(1008 \gev)^2$}
\\ \hline\hline
%%%%%
%%%%%
 {\bf Case II} & {\footnotesize $A_t=-0.2~m_0$}  & {\footnotesize $\tb=50$}~  & {\footnotesize ${\bf \Delta_{m_0^2}=16}$}  
\\ %\hline
 {\footnotesize ${\bf m_0^2}$} & {\footnotesize $({\bf 5.5} \tev)^2$}  & {\footnotesize $({\bf 4.5} \tev)^2$}  & {\footnotesize $({\bf 3.5} \tev)^2$}  
\\ \hline
 {\footnotesize $m_{q_3}^2(t_T)$} &  {\footnotesize $(4376 \gev)^2$}  & {\footnotesize $(3563 \gev)^2$}  & {\footnotesize $(2752 \gev)^2$} 
\\
 {\footnotesize $m_{u^c_3}^2(t_T)$} &  {\footnotesize $(3798 \gev)^2$}  & {\footnotesize $(3106 \gev)^2$}  & {\footnotesize $(2413 \gev)^2$} 
\\
 {\footnotesize ${\bf m_{h_u}^2(t_T)}$} & {\footnotesize $({\bf 539} \gev)^2$}  & {\footnotesize $({\bf 361} \gev)^2$}  & {\footnotesize $-({\bf 44} \gev)^2$}  
\\
 {\footnotesize $m_{h_d}^2(t_T)$} &  {\footnotesize $(2053 \gev)^2$}  & {\footnotesize $(1565 \gev)^2$}  & {\footnotesize $(1046 \gev)^2$}
\\ \hline\hline
%%%%%
%%%%%
 {\bf Case III} & {\footnotesize $A_t=-0.5~m_0$}  & {\footnotesize $\tb=50$}~  & {\footnotesize ${\bf \Delta_{m_0^2}=9}$}  
\\ %\hline
 {\footnotesize ${\bf m_0^2}$} & {\footnotesize $({\bf 5.5} \tev)^2$}  & {\footnotesize $({\bf 4.5} \tev)^2$}  & {\footnotesize $({\bf 3.5} \tev)^2$}  
\\ \hline
 {\footnotesize $m_{q_3}^2(t_T)$} &  {\footnotesize $(4284 \gev)^2$}  & {\footnotesize $(3532 \gev)^2$}  & {\footnotesize $(2630 \gev)^2$} 
\\
 {\footnotesize $m_{u^c_3}^2(t_T)$} &  {\footnotesize $(3755 \gev)^2$}  & {\footnotesize $(3088 \gev)^2$}  & {\footnotesize $(2373 \gev)^2$} 
\\
 {\footnotesize ${\bf m_{h_u}^2(t_T)}$} & {\footnotesize $-({\bf 363} \gev)^2$}  & {\footnotesize $-({\bf 41} \gev)^2$}  & {\footnotesize $-({\bf 546} \gev)^2$}  
\\
 {\footnotesize $m_{h_d}^2(t_T)$} &  {\footnotesize $(1447 \gev)^2$}  & {\footnotesize $(1359 \gev)^2$}  & {\footnotesize $-(950 \gev)^2$}
\\ \hline\hline
%%%%%
%%%%%
 {\bf Case IV} & {\footnotesize $A_t=0$}  & {\footnotesize $\tb=25$}~  & {\footnotesize ${\bf \Delta_{m_0^2}=57}$}  
\\ %\hline
 {\footnotesize ${\bf m_0^2}$} & {\footnotesize $({\bf 5.5} \tev)^2$}  & {\footnotesize $({\bf 4.5} \tev)^2$}  & {\footnotesize $({\bf 3.5} \tev)^2$}  
\\ \hline
 {\footnotesize $m_{q_3}^2(t_T)$} &  {\footnotesize $(4915 \gev)^2$}  & {\footnotesize $(4025 \gev)^2$}  & {\footnotesize $(3134 \gev)^2$} 
\\
 {\footnotesize $m_{u^c_3}^2(t_T)$} &  {\footnotesize $(3770 \gev)^2$}  & {\footnotesize $(3086 \gev)^2$}  & {\footnotesize $(2400 \gev)^2$} 
\\
 {\footnotesize ${\bf m_{h_u}^2(t_T)}$} & {\footnotesize $({\bf 152} \gev)^2$}  & {\footnotesize $-({\bf 220} \gev)^2$}  & {\footnotesize $-({\bf 293} \gev)^2$}  
\\
 {\footnotesize $m_{h_d}^2(t_T)$} &  {\footnotesize $(5057 \gev)^2$}  & {\footnotesize $(4136 \gev)^2$}  & {\footnotesize $(3215 \gev)^2$}
\end{tabular}
\end{center}\caption{Soft squared masses of the stops and Higgs bosons at $t=t_T\approx 8.2$ ($Q_T=3.5 \tev$) for various trial $m_0^2$s 
%, i.e., $m_0^2=(5 \tev)^2$, $(4 \tev)^2$, and $(3 \tev)^2$,  
when the messenger scale is $Q_M\approx 1.3\times 10^{16} \gev$. 
%$y_S\langle S\rangle$ is  $t_M=36.1$ ($Q_M=5\times 10^{15} \gev$). 
$\Delta_{m_0^2}$ indicates the fine-tuning measure for $m_0^2$ around ${(4.5 \tev)^2}$ 
for each case. 
%Wide ranges of $\tb$ and (negative) $A_t$ admit $\Delta_{m_0^2}<100$, if the mGrM of SUSY breaking is assisted GUT scale gauge mediation effects. 
%Case I provides the minimum of $\Delta_{m_0^2}$. 
%$\tb=50$ and $m_{1/2}= A_0=1 \tev$ with $\alpha_{G}=1/24$. 
%The left [right] four columns correspond to the results of $\{y_{NI}^2=0.8,~\widetilde{m}^2=(15 \tev)^2\}$ [$\{y_{NI}^2=1.0,~\widetilde{m}^2=(20 \tev)^2\}$]. 
$m_{h_u}^2$s further decrease to be negative below $t=t_T$.    
The above mass spectra are generated using SOFTSUSY  \cite{softsusy}.
}\label{tab:data}
\end{table}
%
%%%%%%%%%%%%%%%%%%%%%%%%%%%%%%%%%%%%%%%%%%%%%%%%%%%%%%%%%%%%%%%%%%
%%%%%%%%%%%%%%%%%%%%%%%%%%%%%%%%%%%%%%%%%%%%%%%%%%%%%%%%%%%%%%%%%%

Table~\ref{tab:data} lists the soft squared masses at $t=t_T$ 
for the LH and RH stops, and the two MSSM Higgs bosons under the various $m_0^2$s, 
when the messenger scale is $Q_M\approx 1.3\times 10^{16} \gev$, and $\tb$ is $50$ or $25$. 
We can see the changes of $m_{h_u^2}^2$ are quite small [$\ll (550 \gev)^2$] under the changes of $m_0^2$ [$(5.5 \tev)^2$--$(3.5 \tev)^2$] unlike the other soft squared masses, 
because $m_{h_u}^2$ is well-focused at $t=t_T$.   
Case I-IV yield again the same low energy gauginos masses as \eq{lowM_a}, 
because \eq{M_a} is valid at low energy, independent of $A_t$ and $\tb$.  
%\dis{
%M_{3,2,1}\approx \{2.3 \tev , ~912 \gev , ~504 \gev \} 
%} 
%for  
%for $m_0^2=(4 \tev)^2$. 
%They are the predictions of this model, 
%since $m_0^2=(4 \tev)^2$ selected for the $126 \gev$ Higgs mass as well as $3$-$4 \tev$ stop mass uniquely determines the gaugino masses 
%via \eqs{GGsoft}, (\ref{GM}) and (\ref{RGgauge}). 
%They would be testable at LHC run2. 
$A_t$ at low energy turns out to be around $1 \tev$ or smaller 
for $m_0^2=(4.5 \tev)^2$, 
and so its contribution to the Higgs boson mass is still suppressed.
By \eq{RGsm} $m_{h_u}^2$s further decrease to be negative below $t=t_T$.     
%the low energy values of $m_{h_u}^2$ are estimated as 
%\dis{
%$\{-(592 \gev)^2, -(331 \gev)^2, -(550 \gev)^2, -(458 \gev)^2\}$ 
%} 
%for Case I, II, III, and IV, respectively.  
With \eq{m_Z} $|\mu|$ are determined as 
$\{485 \gev, 392 \gev, 516 \gev, 586 \gev\}$ 
for Case I, II, III, and IV, respectively.
Actually the RG running of $\mu$ is completely separated from other soft parameters. 
Moreover, the generation scale of $\mu$ is quite model-dependent. 
So we don't discuss them here. 
To avoid a potential fine-tuning issue associated with $\mu$, 
however, we confine our discussion to cases of $|\mu|<600 \gev$.

From Table~\ref{tab:data}, we can read the $A_t$ dependence of the fine-tuning measure $\Delta_{m_0^2}$ 
($\equiv \left|\frac{\partial{\rm log}m_Z^2}{\partial {\rm log}m_0^2}\right|
=\left|\frac{m_0^2}{m_Z^2}\frac{\partial m_Z^2}{\partial m_0^2}\right|$ \cite{FTmeasure}) 
around $m_0^2=(4.5 \tev)^2$. 
Case I gives almost the minimum of $\Delta_{m_0^2}$ ($=1$) when $\tb=50$.  
%whereas Case I and III correspond to the boundaries of $\Delta_{m_0^2}\lesssim 100$ with the same $\tb$. 
%[When $A_t/m_0=+0.1$ ($-0.8$), $\Delta_{m_0^2}$ turns out to be $121$ ($107$).]  
On the other hand, $\Delta_{A_t}$ ($=\left|\frac{A_t}{m_Z^2}\frac{\partial m_Z^2}{\partial A_t}\right|$) are $\{0, 10, 118, 0\}$ 
for Case I, II, III, and IV, respectively. 
When $A_t/m_0=+0.1$, $\{\Delta_{m_0^2}, \Delta_{A_t}, |\mu|\}$ turn out to be about $\{22, 33, 569 \gev\}$. 
%$\Delta_{A_t}$ drastically increases below $A_t/m_0=-0.5$: 
%$\Delta_{A_t}$ is $119$ for $A_t/m_0=-0.6$, while it is $57$ for $A_t/m_0=-0.5$. 
Therefore, we can conclude the parameter range 
\dis{
-0.5 ~<~  A_t/m_0 ~\lesssim~ +0.1 
~~~{\rm and}~~ \tb \gtrsim 25  
}  
allows $\{\Delta_{m_0^2}, \Delta_{A_t}\}$ and $|\mu|$ 
to be smaller than $100$ 
and $600 \gev$, respectively. 
%Case IV is the case of $\tb=25$ when $A_t/m_0=0$. It gives$\{\Delta_{m_0^2}, \Delta_{A_t}\}=\{57,0\}$.
%, while Case I-III are for $\tb=50$.  
We see that a larger $\tb$ would be preferred for a smaller $\Delta_{m_0^2}$. 
It is basically because $m_{h_d}^2$ is not focused unlike $m_{h_u}^2$, 
even though it also contributes to $m_Z^2$ as seen in \eq{m_Z}. 
Actually $\tb=50$ is easily obtained e.g. from the minimal SO(10) GUT \cite{GUT}.

%%%%%%
%%%%%%  
% XXX
%%%%%%
%%%%%%

In the above cases, the sleptons and sbottom 
(superpartners of the leptons and b-quark) 
turn out to be quite heavier than $3 \tev$. 
The first two generations of SUSY particles would be much heavier than them.   
%
%The implication of this spectrum on dark matter cosmology  
%will be discussed elsewhere \cite{on-going}. 
%  
%
%For $M_1<|\mu|$ ($M_1>|\mu|$), 
Hence, the bino is the lightest superparticle (LSP). 
%For $M_1<|\mu|$, 
To avoid overclose of the bino dark matter in the Universe, 
some entropy production \cite{DM1} or 
%On the other hand, for $M_1>|\mu|$, the Higgsino dark matter needs to be supplemented with other components such as the axion \cite{DM2}.   
other lighter dark matter such as the axino and axion is needed \cite{DM2}. 
Further numerical analyses on the parameter space will be found 
in other literatures \cite{on-going}.

In conclusion, we have noticed that a FP of $m_{h_u}^2$ appears at $3$-$4 \tev$, 
when the mGrM and mGgM effects are combined at the GUT scale 
for a common SUSY breaking source parametrized with $m_0$, 
and the gaugino masses are dominantly generated by the mGgM effects. 
Even for a $3$-$4 \tev$ stop mass explaining the $126 \gev$ Higgs mass, 
thus, the fine-tuning measures significantly decrease well below $100$ 
for $-0.5< A_t/m_0\lesssim +0.1$ and $\tb \gtrsim 25$ 
in the minimal mixed mediation.   
In this range, $|\mu|$ is smaller than $600 \gev$.     
The expected gluino mass is about $1.7 \tev$, 
which could readily be tested at LHC run2.

%\vspace{5mm}

\acknowledgments

%B.K. thanks Department of Physics and Astronomy in Rutgers University for the hospitality during his visit to Rutgers University. 
{\it Acknowledgments} B.K. thanks Doyoun Kim for useful discussions. 
B.K. is supported by 
%Basic Science Research Program through 
the National Research Foundation of Korea (NRF) funded by the Ministry of Education, Grant No. 2013R1A1A2006904, 
and also in part 
%%B.K. acknowledges the support 
by Korea Institute for Advanced Study (KIAS) grant funded by the Korean government.

%%%%%%%%%%%%%%%%%%%%%%%%%%%%%%%%%%%%%%%%%%%%%%%%%%%%%%%%
%%%%%%%%%%%%%%%%%%%%%%%%%%%%%%%%%%%%%%%%%%%%%%%%%%%%%%%%%

\end{document}